\documentclass[superscriptaddress,showpacs,prl,reprint]{revtex4-1}

\usepackage{hyperref}
\usepackage{graphicx}
\usepackage{color}
\usepackage{amsbsy}
\usepackage{pstricks}
\usepackage{multirow}
\usepackage{amsmath}
\usepackage{lipsum}

\begin{document}

\title{Spin-orbit-induced circulating currents in a semiconductor nanostructure}

\date{\today}

\author{J. van Bree}
\email{j.v.bree@tue.nl}
\author{A. Yu. Silov}
\author{P. M. Koenraad}
\affiliation{PSN, COBRA Research Institute, Eindhoven University of Technology, 5600 MB Eindhoven, The Netherlands}
\author{M. E. Flatt{\'e}}
\email{michael\_flatte@mailaps.org}
\affiliation{PSN, COBRA Research Institute, Eindhoven University of Technology, 5600 MB Eindhoven, The Netherlands}
\affiliation{Department of Physics and Astronomy and Optical Science and Technology Center, University of Iowa, Iowa City, Iowa 52242, USA}


\begin{abstract}
Circulating orbital currents produced by the spin-orbit interaction for a single electron spin in a quantum dot are explicitly evaluated at zero magnetic field, along with their effect on the total magnetic moment (spin and orbital) of the electron spin. The currents are dominated by coherent superpositions of the conduction and valence envelope functions of the electronic state, are smoothly varying within the quantum dot, and are peaked roughly halfway between the dot center and edge. Thus the spatial structure of the spin contribution to the magnetic moment (which is peaked at the dot center) differs greatly from the spatial structure of the orbital contribution. Even when the spin and orbital magnetic moments cancel (for $g=0$) the spin can interact strongly with local magnetic fields, {\it e.g.} from other spins, which has implications for spin lifetimes and spin manipulation.
\end{abstract}

\maketitle

Spin-correlated orbital currents dramatically modify the  magnetic moment ${\boldsymbol \mu}$ of an electron spin in many semiconductors, often enhancing ${\boldsymbol \mu}$ by an order of magnitude over its free-electron value\cite{Roth1959,Yafet1963}. This modified magnetic moment also controls the spin dynamics in nanostructures, and is usually parametrized in the literature as a shape, size and composition dependent $g$ tensor defined by ${\boldsymbol \mu} = {\mathbf g}\cdot {\bf S}$, where ${\bf S}$ represents the electron spin\cite{Kiselev1998,Rodina2003,Schrier2003,Nenashev2003,Prado2003,Prado2004,Krasny2001,Pryor2006b,Pryor2007,Nakaoka2007,Pingenot2008,Andlauer2009,Roloff2010,Bree2012}. Despite the central nature of $g$ tensors to high-speed spin manipulation\cite{Nakaoka2007,Pingenot2008,Andlauer2009,De2009,Roloff2010,Pingenot2011}, spin lifetimes\cite{Florescu2006,Ban2012}, and quantum computation\cite{Loss1998}, the spatial structure of the spin-correlated orbital currents that determine these $g$ tensors has not been investigated. The resulting spatial structure of the magnetic moment ${\boldsymbol \mu}({\bf r})$, which has been neglected up to now, would significantly affect the interaction of confined electron spins with magnetic fields that vary rapidly in space, such as from nearby spins. The most natural assumption, that ${\boldsymbol \mu}({\bf r}) = {\boldsymbol \mu}_{\text{eff}}|\Psi({\bf r})|^2$,
where $\Psi({\bf r})$ is the wave function of the ground-state electron, is incorrect. 
Recently the nature of spin-correlated orbital currents has been investigated in magnetic metals and insulators in a spatial formulation that identifies itinerant circulating currents at the edges of materials\cite{Thonhauser2005,Lopez2012}, by constructing the orbital contribution to the magnetic moment originating from each unit cell in the material. The itinerant circulating currents of Refs.~\onlinecite{Thonhauser2005,Lopez2012}, however, are relatively small by comparison with the large itinerant currents that can arise for carriers in semiconductors and semimetals, as was first shown in the diamagnetic response of bismuth\cite{Ehrenfest1925,Ehrenfest1929,Yafet1963}. Spin-correlated orbital currents also play a key role in the fundamental understanding and phenomenology of the quantum spin Hall effect\cite{Kane2005,Bernevig2006b}.

Here we calculate the spatial distribution of spin-correlated orbital currents for the lowest-energy electron spin states of a quantum dot {\it at zero magnetic field}, and show that 
itinerant currents are extended throughout the quantum dot, peaking about midway out from the center of the dot. We assume that a single electron resides in the quantum dot with an oriented spin, {\it e.g.} through spin injection or optical excitation, and that no magnetic field is applied; the resulting circulating currents thus are not due to an orbital response to an applied magnetic field.  For spherical dots with hard-wall boundary conditions the electronic states are obtained within an analytically-solvable envelope-function formalism; results for nanowire quantum dots and quantum-well quantum dots are also presented. In each case the magnetic moment is primarily due to itinerant currents originating  from coherent superpositions of conduction and valence envelope functions, rather than from magnetic moments associated with the Wannier functions of each unit cell (in contrast to Refs.~\cite{Thonhauser2005,Lopez2012}). In this we find a spin-orbit analogue to known features of the spinless orbital angular momentum in a magnetic field, which was shown long ago to be due almost entirely to itinerant currents\cite{Ehrenfest1925,Ehrenfest1929,Yafet1963}.  In the limit of large dot size, approaching the bulk limit, the effective range of the orbital currents is set by the de Broglie wavelength. The resulting correct form of ${\boldsymbol \mu}({\bf r})$ differs greatly from  ${\boldsymbol \mu}_{\text{eff}}|\Psi({\bf r})|^2$. For example, the spin of a quantum dot with $g=0$ will still evince a local magnetic moment that could interact with localized magnetic systems such as ferromagnets\cite{Pioro2008} or nuclear moments through the hyperfine interaction, as well as a quadrupolar magnetic moment
that could couple to nearby spins. Orbital angular momentum quenching\cite{VanVleck1932} in quantum dots\cite{Pryor2006b} thus consists of reducing the amplitude of the orbital magnetic moment generated by this itinerant current.

\begin{figure}[t]
\begin{center}
\includegraphics[width=0.8\columnwidth]{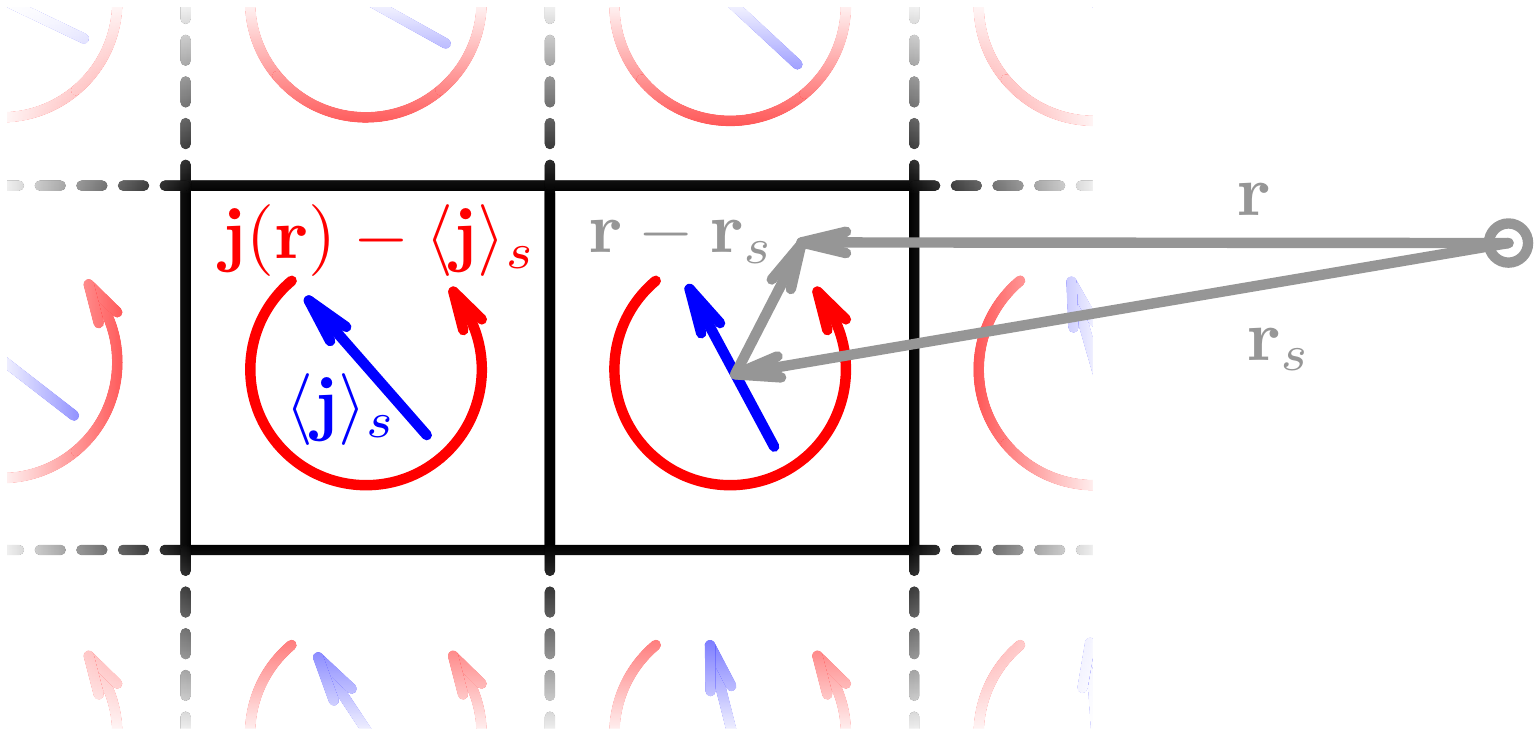}
\caption{The orbital current within a unit cell can be split into an itinerant contribution $\langle{\bf j}\rangle_s$, and a localized contribution ${\bf j}({\bf r}) - \langle{\bf j}\rangle_s$. Vector ${\bf r}_s$ points to the center of unit cell $s$.}
\label{fig:IC_LC}
\end{center}
\end{figure}

The total magnetic moment ${\boldsymbol \mu}$ is the sum of the spin magnetic moment ${\boldsymbol \mu}_{\text{spin}}$ and the orbital magnetic moment ${\boldsymbol \mu}_{\text{orb}}$.  The orbital magnetic moment ${\boldsymbol \mu}_{\text{orb}}$ of a stationary state $\Psi\left({\bf r}\right)$ occupying a volume $V$ is related to the orbital current density ${\bf j}\left({\bf r}\right)$ \cite{Jackson1998b}, by
\begin{eqnarray}
{\boldsymbol \mu}_{\text{orb}} &=& \frac{1}{2}\int_V{\bf r}\times{\bf j}\left({\bf r}\right)d{\bf r} = \frac{1}{2} \sum_s \int_{V_s} {\bf r} \times {\bf j}({\bf r}) d{\bf r}, \label{eq:mu}
\end{eqnarray}
where we have considered the moment as a summation of moments arising from currents {\bf j}({\bf r}) flowing in each of $s$ unit cells having volume $V_s$.
For the ground state of an electron in a quantum dot, with maximal spin along an axis, the total spatially-integrated current must vanish ({\it i.e.} a current loop). To find the scale of these current loops we decompose the orbital current into an itinerant current that flows into or out of a unit cell, $\langle{\bf j}\rangle_s$, and a localized current whose average over the unit cell vanishes, ${\bf j}({\bf r}) - \langle{\bf j}\rangle_s$; the distinction between these is shown graphically in Fig.~\ref{fig:IC_LC}. In semiconductor quantum dots the orbital magnetic moment from the itinerant current dominates the total orbital magnetic moment, and varies slowly on the scale of unit cells.

The orbital magnetic moment can then be expressed\cite{Thonhauser2005} as:
\begin{eqnarray}
{\boldsymbol \mu}_{\text{orb}} &=& \frac{1}{2} \sum_s \Big[V_s {\bf r}_s \times \langle {\bf j} \rangle_s + \int_{V_s} ({\bf r} - {\bf r}_s) \times \left\{{\bf j}({\bf r}) - \langle {\bf j} \rangle_s\right\} d{\bf r}\Big]\nonumber
\end{eqnarray}
where ${\bf r}_s$ is the vector pointing to unit cell $s$. The first term is the orbital moment due to itinerant currents, whereas the second term is the sum of orbital moments due to a (circulating) current localized within each unit cell. For an isolated atom the first term is zero.

Within the envelope function approximation, which is appropriate for quantum dots much larger than a unit cell of the constituent material, the wave function $\Psi({\bf r})$ is the product of a Bloch state $u_i({\bf r})$ of band~$i$, and a  spatially slowly varying envelope function $F_i\left({\bf r}\right)$ (approximately constant in a unit cell),
\begin{eqnarray}
\Psi\left({\bf r}\right)=\sum_i F_i\left({\bf r}\right) u_i({\bf r}),
\end{eqnarray}
and   ${\bf j}({\bf r}) = (e\hbar/m_0)\text{Im}\left\{\Psi^*\left({\bf r}\right)\nabla\Psi\left({\bf r}\right)\right\} $ \cite{Messiah1961} becomes 
\begin{eqnarray}
{\bf j}\left({\bf r}\right) &=&  \frac{e\hbar}{m_0} \sum_{i,j} \text{Im}\left\{u_i^*({\bf r}) u_j({\bf r}) F_i^*\left({\bf r}\right)\nabla F_j\left({\bf r}\right) \right. \nonumber \\*
 & &\qquad\qquad+ \left. F_i^*\left({\bf r}\right) F_j\left({\bf r}\right) u_i^*({\bf r}) \nabla u_j({\bf r})\right\}. \label{eq:j_kp}
\end{eqnarray}
The first term contains the velocity generated by the envelope wave function, whereas for the second term the velocity comes from the Bloch functions. Contributions to the moment come either from the Bloch velocity term (BV) or envelope velocity term (EV) of Eq.~\eqref{eq:j_kp}, and originate either from the cell-averaged current $ \langle {\bf j} \rangle_s$ (IC) or the current within the unit cell (LC). For concreteness we consider the orbital moment for an electron in a nanostructure whose Bloch functions $u_i({\bf r})$ are eigenstates of parity. 
Defining ${\boldsymbol \mu}_{\text{orb}} = {\boldsymbol \mu}_{\rm EV} + {\boldsymbol \mu}_{\rm BV}$ and ${\boldsymbol \mu}_{\rm BV} = {\boldsymbol \mu}_{\rm BV,IC}+ {\boldsymbol \mu}_{\rm BV,LC}$, and simplifying using the symmetries of the Bloch functions, we find the orbital magnetic moment densities
\begin{eqnarray}
{\boldsymbol \mu}_{\rm BV,IC}({\bf r}_s) &=& \frac{e \hbar}{2 m_0} \sum_{i,j} \text{Im}\left\{F_i^*({\bf r}_s)F_j({\bf r}_s) ({\bf r}_s \times \langle u_i | \nabla | u_j \rangle)\right\},\nonumber \\*
{\boldsymbol \mu}_{\rm BV,LC}({\bf r}_s) &=& \frac{e \hbar}{2 m_0} \sum_{i,j} \text{Im}\left\{F_i^*({\bf r}_s)F_j({\bf r}_s) \langle u_i|{\bf L}_{\text{B}}|u_j\rangle\right\},\nonumber \\*
{\boldsymbol \mu}_{\rm EV}({\bf r}_s) &=& \frac{e \hbar}{2 m_0} \sum_{i\ne j} \text{Im}\left\{
\langle u_i | {\bf r} | u_j \rangle\times  F_i^*({\bf r}_s)\nabla F_j({\bf r}_s)\right\} \nonumber \\*
&&+\frac{e \hbar}{2 m_0} \sum_{i} F_i^*({\bf r}_s){\bf L}_{\text{E}} F_i({\bf r}_s),\nonumber 
\end{eqnarray}
where ${\bf L}_{\text{B}}$ and ${\bf L}_{\text{E}}$ are the angular momentum operators acting on the Bloch and envelope functions respectively. Studies of optical matrix elements, whose values also depend on the current appearing in Eq.~(\ref{eq:j_kp}), have established that matrix elements of the Bloch velocity exceed those of the envelope velocity by approximately the ratio of the nanostructure linear size to the unit cell size\cite{Johnson1987}.  This ratio is $\gtrsim15$ for realistic parameters, and thus, as will be evident below, ${\boldsymbol \mu}_{\rm EV} \ll {\boldsymbol \mu}_{\rm BV}$ for the range of validity of the envelope function approximation.  Furthermore, as $ \langle u_i|{\bf L}_{\text{B}}|u_j\rangle$ --- the angular momentum of the Bloch function --- does not exceed $1$, the dominant contribution to the orbital magnetic moment must be from ${\boldsymbol \mu}_{\rm BV,IC}\gg {\boldsymbol \mu}_{\rm BV,LC}$.

\begin{figure}[t]
\begin{center}
\includegraphics[width=\columnwidth]{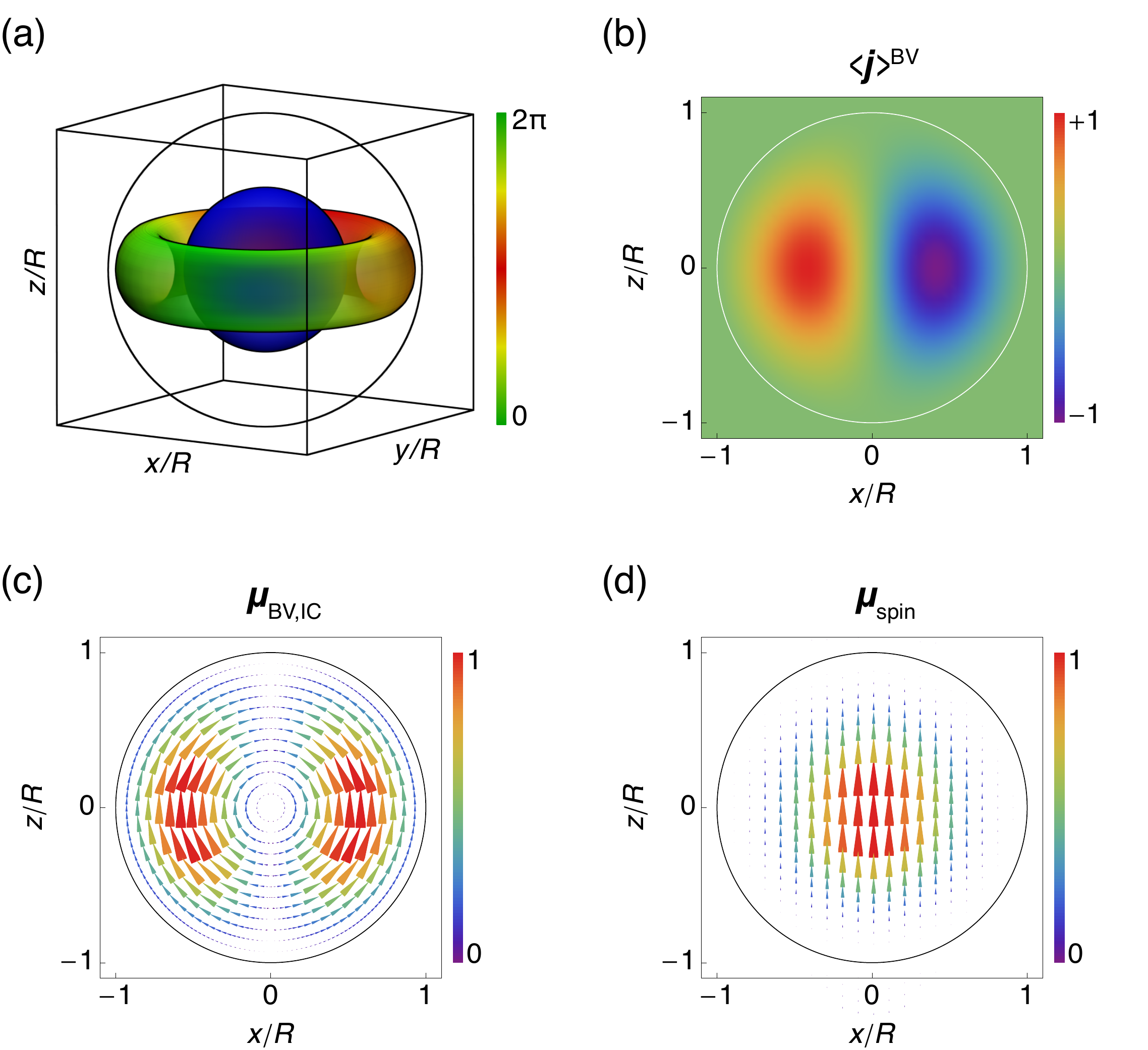}
\caption{(a) An illustration of the contours of constant probability of the conduction band (blue) and valence band (green-red) envelope functions. The valence band envelope function is colored according to its phase. (b) The normalized magnitude of $\langle{\bf j}\rangle^{\text{BV}}$ in the ${\bf e}_y$-direction. (c) The normalized dominant contribution to the orbital magnetic moment ${\boldsymbol \mu}_{\text{BV,IC}}$. (d) The normalized spin magnetic moment ${\boldsymbol \mu}_{\text{spin}}$. (b-d) are $xz$-cross-sections, where the boundary of the sphere is represented by the white/black circle.}
\label{fig:jmu}
\end{center}
\end{figure}

We now consider the origin of the orbital moment for an electron in the lowest conduction state of a quantum dot. 
The minimal set of Bloch states are two conduction $s$ states and six valence $p$ states (an eight-band ${\bf k}\cdot{\bf p}$ model for the semiconductor, as in Ref.~\onlinecite{Vahala1990,Sercel1990}). To avoid complications from other spin-dependent effects we neglect the zero-field spin splittings of conduction-band states, and other inversion-asymmetric effects, in III-V semiconductors.
For $i$ labeling the conduction band $F_i\left({\bf r}\right)$ is  $s$-like, whereas for $j$ labeling a valence state $F_j\left({\bf r}\right)$ is $p$-like. The spatial distribution of ${\bf j}\left({\bf r}\right)$, dominated by the product $F_i^*\left({\bf r}\right) F_j\left({\bf r}\right)$, will therefore peak between the center and edge of the nanostructure. This spatial dependence is significantly different from that of  $|\Psi({\bf r})|^2$. For a stationary state, the divergence of ${\bf j}\left({\bf r}\right)$ is zero and the current must flow along a closed surface within the nanostructure. Therefore ${\bf j}\left({\bf r}\right)$ resembles a current loop, see Fig.~\ref{fig:jmu}(a). If the nanostructure is very large the de Broglie wavelength, or the Bohr radius of dopants, will set the length scale associated with the current loop. As the nanostructure gets smaller, quantum confinement  quenches the current loop and modifies the electron $g$-factor.

The spatial structure of the spin magnetic moment is completely different from that of the orbital magnetic moment of the electron spin. In the non-relativistic limit of the Dirac equation\cite{Messiah1961} the spin magnetic moment ${\boldsymbol \mu}_{\text{spin}}$ is 
\begin{eqnarray}
{\boldsymbol \mu}_{\text{spin}} &=& \frac{e \hbar}{2 m_0} \int_V \Psi^*({\bf r}){\boldsymbol \sigma}\Psi({\bf r}) d{\bf r} \nonumber \\*
&=& \frac{e \hbar}{2 m_0} \sum_{s,i,j} F_i^*({\bf r}_s) F_j({\bf r}_s) \langle u_i | {\boldsymbol \sigma} | u_j \rangle,
\end{eqnarray}
where ${\boldsymbol \sigma}$ is the Paul vector. The spatial distribution of ${\boldsymbol \mu}_{\text{spin}}$ is determined by $F_i^*({\bf r}_s)F_j({\bf r}_s)$. If the dominant envelope function is that of the conduction band, then the largest contribution to the  spin magnetic moment will have spatial structure $|F_i({\bf r}_s)|^2$, where $F_i\left({\bf r}\right)$ is $s$-like. Thus no current loop structure exists for the spin moment.

The dependence on dot radius $R$ of the electron ground state orbital magnetic moment for a quantum dot provides a concrete demonstration of these features. Here we summarize the calculation for spherical dots with hard-wall boundary conditions; those for nanowire quantum dots or quantum-well quantum dots are described in Supplementary Material\cite{suppl}. The Hamiltonian ${\cal H}$ commutes with the total angular momentum ${\bf F}={\bf L}_{\text{E}}+{\bf L}_{\text{B}}+{\bf s} = {\bf L}_{\text{E}} + {\bf J}$ (${\bf s}$ is the spin moment and ${\bf J}$ the total magnetic moment of the Bloch function), so the Hamiltonian is block diagonal in a basis of $|F, F_z\rangle$ \cite{Vahala1990,Sercel1990},
\begin{eqnarray}
{\cal H} = \sum_{F,F_z} {\cal H}_{F,F_z},
\end{eqnarray}
where the  electron ground state of the quantum dot is $|F,F_z\rangle = |\frac{1}{2},\pm\frac{1}{2}\rangle$. As this is a Kramers doublet, it suffices to examine  $|\frac{1}{2},+\frac{1}{2}\rangle$ to understand the angular-momentum structure, as $|\frac{1}{2},-\frac{1}{2}\rangle$ is simply the time-reverse of $|\frac{1}{2},+\frac{1}{2}\rangle$. An eight band ${\bf k}\cdot{\bf p}$-model for ${\cal H}$ is analytically solvable, and  only three spherical basis states $|F,F_z;J,L_{\text {E}}\rangle$ contribute to $|\frac{1}{2},+\frac{1}{2}\rangle$,
\begin{eqnarray}
|\tfrac{1}{2},+\tfrac{1}{2}\rangle = \frac{|\tfrac{1}{2},\tfrac{1}{2};\tfrac{1}{2},0\rangle + \alpha |\tfrac{1}{2},\tfrac{1}{2};\tfrac{3}{2},1\rangle + \beta |\tfrac{1}{2},\tfrac{1}{2};\tfrac{1}{2},1\rangle}{(1+|\alpha|^2+|\beta|^2)^{1/2}}, \nonumber
\end{eqnarray}
which follows from the rules for adding angular momenta and the parity of the effective mass equations~\cite{Shechter1962}. The problem is now  analytically solvable; diagonalizing the Hamiltonian yields $\alpha=\alpha(R)$ and $\beta=\beta(R)$. These coefficients represent the degree of intermixing of valence band states into the electron ground state. They are small and the ground state is dominated by the conduction band contribution $|\tfrac{1}{2},\tfrac{1}{2};\tfrac{1}{2},0\rangle$. See Supplemental Material\cite{suppl} for explicit expressions for $\alpha$ and $\beta$.

\begin{figure}[t]
\includegraphics[width=\columnwidth]{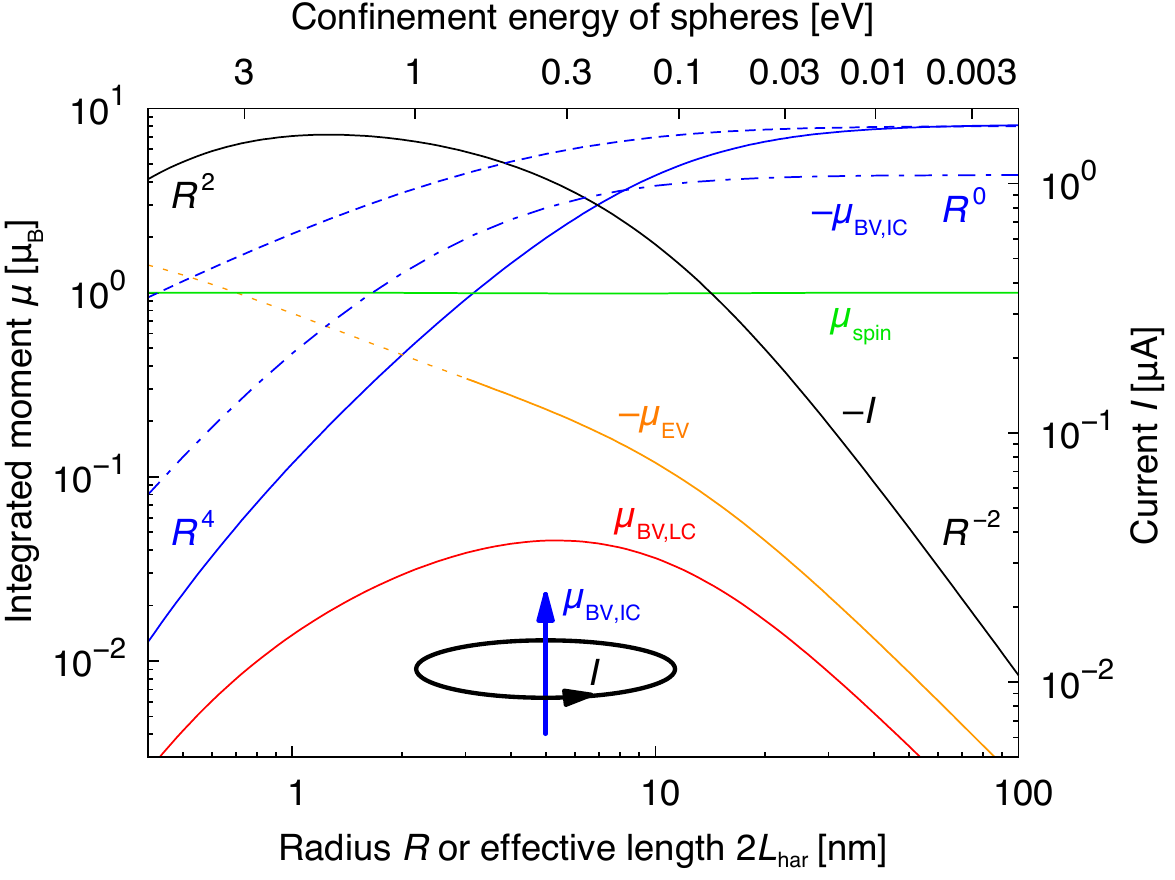}
\caption{Contributions to the orbital moment ${\boldsymbol \mu}_{\text{orb}}$, the spin moment ${\boldsymbol \mu}_{\text{spin}}$ and current $I$ of an InAs sphere with a hard wall boundary, as a function of the radius $R$ and the confinement energy. Also included are InAs quantum-well based quantum dots (dot-dashed line)\cite{Hanson2007} and nanowire quantum dots (dashed line)\cite{Csonka2008}.}
\label{fig:mu}
\end{figure}

The cell-averaged current originating from the Bloch velocity, $\langle{\bf j}\rangle^{\text{BV}}$, is the source of the dominant orbital moment ${\boldsymbol \mu}_{\text{BV,IC}}$. From the wave function of the $|\frac{1}{2},+\frac{1}{2}\rangle$ state,  using Eq.~(\ref{eq:j_kp}), we calculate $\langle{\bf j}\rangle^{\text{BV}}$ to be
\begin{eqnarray}
\langle{\bf j}\rangle^{\text{BV}} = -\frac{e P_0}{2\sqrt{6} \pi \hbar}\frac{\text{Im}\big\{\alpha-\sqrt{2}\beta\big\}}{1+|\alpha|^2+|\beta|^2} j_0(kr)j_1(kr) \sin(\theta) {\bf e}_{\phi} \nonumber,
\end{eqnarray}
where $j_l(kr)$ is the $l^{\text{th}}$ spherical Bessel function, $k$ the spherical wave number, and $P_0$ the Kane matrix element. As $\langle{\bf j}\rangle^{\text{BV}}$ only has an ${\bf e}_{\phi}$-component it  suffices to show the magnitude of $\langle{\bf j}\rangle^{\text{BV}}$ in the ${\bf e}_y$-direction in an $xz$-cross-section, as shown in Fig.~\ref{fig:jmu}(b). $\langle{\bf j}\rangle^{\text{BV}}$ is proportional to the coherent product of the conduction band and valence band envelope functions, $j_0(kr)j_1(kr)$. The current density therefore peaks roughly at $R/2$ and strongly resembles a classical current loop circulating in the $xy$-plane. The coefficients $\alpha$ and $\beta$ depend on the spin-orbit coupling $\Delta$, and  $\text{Im}\{\alpha-\sqrt{2}\beta\} \propto \Delta$, demonstrating directly the spin-correlated nature of $\langle{\bf j}\rangle^{\text{BV}}$. As expected from the transformation properties of currents and spins under time reversal, $\langle{\bf j}\rangle^{\text{BV}}$ circulates in the opposite direction for the time reversed state $|\tfrac{1}{2},-\tfrac{1}{2}\rangle$.

The spatial structure of the dominant orbital magnetic moment ${\boldsymbol \mu}_{\text{BV,IC}}$ mimics the spatial distribution of the current and is shown in Fig.~\ref{fig:jmu}(c). In contrast, the spatial structure of the spin magnetic moment, shown in Fig.~\ref{fig:jmu}(d), mimics the probability density of the electron's wave function and differs completely from the spatial structure of the orbital moment.

The dependence on $R$ of the circulating orbital currents and the resulting contributions to the orbital magnetic moment are plotted in Fig.~\ref{fig:mu} for an InAs sphere. For large $R$ the magnetic moment ${\boldsymbol \mu}$ approaches the bulk value $\sim 8 \mu_\text{B}$. On the other hand, the magnetic moment approaches zero as $R$ gets smaller. This dependence exhibits the orbital momentum quenching described in Ref.~\onlinecite{Pryor2006b}. For the entire range of values ($R>3$~nm) where the envelope function approximation is valid, ${\boldsymbol \mu}_{\text{BV,IC}}$ is a factor of five or more larger than the other contributions to the orbital magnetic moment,  ${\boldsymbol \mu}_{\text{BV,LC}}$ and ${\boldsymbol \mu}_{\text{EV}}$.
The dominant contribution to the spin moment comes from the conduction band contribution $|\tfrac{1}{2},\tfrac{1}{2};\tfrac{1}{2},0\rangle$, and the calculated spin moment density peaks at the center of the quantum dot, as shown in Fig.~\ref{fig:jmu}(d). The valence band contributions are negligible, since the integrated spin moment, shown in Fig.~\ref{fig:mu}, is within 1\% of one Bohr magneton. In the limit of $R\rightarrow\infty$ our analytic expression for ${\boldsymbol \mu}$ is  identical to Roth's formula\cite{Roth1959}. 

In order to understand the saturating behavior ${\boldsymbol \mu}_{\text{BV,IC}}$ for large $R$, and the orbital angular momentum quenching for small $R$ in more detail, the magnetic moment originating from a current loop provides insight. The magnetic moment of  a loop carrying a current $I$ is
\begin{eqnarray}
\mu_{\text{loop}} = \pi I R^2.
\end{eqnarray}
Quantum confinement restricts the radius $R$ of the loop and  therefore quenches the magnetic moment.  The radius dependence of the current, 
\begin{eqnarray}
I = \int \langle{\bf j}\rangle^{\text{BV}} \cdot {\bf n}~d{\bf A} \sim \frac{\text{Im}\big\{\alpha-\sqrt{2}\beta\big\}}{R}, \label{eq:I}
\end{eqnarray}
is plotted for the InAs sphere in Fig.~\ref{fig:mu}. The current is proportional to the product of the amplitudes of the conduction and valence envelope function, and reaches a maximum around $R\sim1$~nm where the confinement energy is $\sim 1.5$~eV. To better understand the radius dependence of $I$ we need to analyze in detail the conduction-valence coupling. The contribution of valence states to the electron ground state depends on $k$ times the momentum matrix element, divided by the energy splitting between conduction and valence states. Within ${\bf k}\cdot{\bf p}$ theory, therefore, the conduction-valence coupling that determines $\alpha$ and $\beta$ in Eq.~\eqref{eq:I} is proportional to $k\sim1/R$. The energy splitting at large $R$ (small $k$) is approximately constant (and equal to the band gap), and therefore $I\propto1/R^2$ so ${\boldsymbol \mu}$ approaches a constant. However, at small $R$ the energy difference depends on 
the free kinetic energy, and along with additional cancellations between $\alpha$ and $\beta$ for small $R$ the limiting behavior as $R\rightarrow 0$ is $I\propto R^2$ so ${\boldsymbol \mu}\propto R^4$. The maximum of the current as a function of dot radius (Fig.~\ref{fig:mu}) is therefore a competition between the band gap and the free kinetic energy, and peaks when the free kinetic energy equals roughly the band gap energy. Note that a similar dependence on free kinetic energy and the band gap influences the electron energy-dependence of the $g$ factor in bulk semiconductors and leads to $g\rightarrow 2$ for large electron energies\cite{Zawadzki1963,Oestreich1996,Tutuc2002}.
 When  the limiting functional dependence of $I$ is inserted directly into the equation for the magnetic moment of a current loop, one immediately obtains the limiting functional dependence of the magnetic moment, justifying the current-loop analogy for interpreting the origin of the orbital magnetic moment in semiconductor nanostructures. 
 
 We also include in Fig.~\ref{fig:mu} results for InAs quantum-well based quantum dots (height 10~nm and lateral harmonic confinement length $L_{\text{har}}$\cite{Hanson2007}) and InAs nanowire quantum dots (radius 40~nm and harmonic confinement height $L_{\text{har}}$\cite{Csonka2008}), showing that these features are quite general. Additional information on the calculations for such dots is available in the Supplementary Material. 
  We note that strain, such as occurs in Stranski-Krastanov dots, will modify the band edges of the constituent materials and hence change $\alpha$ and $\beta$, but the qualitative analysis of these dots will be similar to those presented here.
 
We expect this approach to be applicable to holes in quantum dots as well, although the structure of the circulating currents is much more complex. Whereas the composition of the electron ground state is predominately the product of an $s$-like envelope function and an $s$-like Bloch function, the dominant composition of the hole ground state is the product of an $s$-like envelope function and a $p$-like Bloch function. The greater Bloch function angular momentum leads to complex orbital momentum structure of the hole state wave functions and sensitive dependences of hole state ordering on size and strain\cite{Margapoti2010,Margapoti2012}.

The direct identification of the circulating currents that produce the orbital magnetic moment for an electron spin within a quantum dot has immediate implications for the spin dynamics, intrinsic magnetism, and $g$ tensor structure of quantum dots. Even when the $g=0$ for the electron spin, the difference between the orbital magnetic moment and the spin magnetic moment means the electron spin can couple to localized magnetic fields, such as those originating from nuclear spins (hyperfine interaction), ionic moments, or nanoscale ferromagnetic regions\cite{Pioro2008}. The lack of a substantial contribution to the circulating current from the center of the dot suggests that the $g$ tensor for quantum rings should be very similar to that of quantum disks, which has been observed experimentally but unexplained\cite{Kleemans2009b}. The nature of these spatially-dependent currents should also influence other observables that depend on currents, such as optical matrix elements that influence oscillator strengths for optical transitions (e.g. as in Ref.~\onlinecite{Johnson1987}). 

MEF acknowledges support from an AFOSR MURI. J. v. B., A. Yu. S., and P. M. K. acknowledge financial support by the COBRA Research Institute.

\end{document}